\documentclass[runningheads]{llncs}

\usepackage{amsmath,amssymb,amsfonts}
\usepackage{algorithmic}
\usepackage{graphicx}
\usepackage{textcomp}
\usepackage{xcolor}
\usepackage{subfig}
\usepackage{caption} 
\usepackage{my_style}
\usepackage{listings}
\usepackage[backend=biber,natbib=true,maxbibnames=99]{biblatex}
\providecommand{\keywords}[1]
{
  \small	
  \textbf{\textit{Keywords---}} #1
}

\title{Unsupervised dynamic modeling of medical image transformation}
\author{Niklas~Gunnarsson\inst{1,2}\orcidID{0000-0002-9013-949X} \and Jens~Sj\"olund\inst{2}\orcidID{0000-0002-9099-3522} \and Peter Kimstrand\inst{2}\orcidID{0000-0001-9667-5595} \and 
Thomas~B.~Sch\"on\inst{1}\orcidID{0000-0001-5183-234X}}
\authorrunning{N. Gunnarsson et al.}
\institute{Department of Information Technology, Uppsala University, Sweden
\email{\{firstname\}.\{surname\}@it.uu.se}\\
 \and
 Elekta Instrument AB, Stockholm, Sweden \\ \email{\{firstname\}.\{surname\}@elekta.com}}

\begin{document}
\maketitle
\noindent \textbf{Please cite this version:} \\
\fullcite{gunnarsson2022unsupervised} \\
\begin{lstlisting}
@inproceedings{gunnarsson2022unsupervised,
  title={Unsupervised dynamic modeling of medical 
  image transformations},
  author={Gunnarsson, Niklas and Sj{\"o}lund, Jens 
  and Kimstrand, Peter and Sch{\"o}n, Thomas B},
  booktitle={2022 25th International Conference on 
  Information Fusion (FUSION)},
  pages={01--07},
  year={2022},
  organization={IEEE}
}
\end{lstlisting}
\newpage
\title{Unsupervised dynamic modeling of medical image transformations}
\author{Niklas~Gunnarsson\inst{1,2}\orcidID{0000-0002-9013-949X} \and Jens~Sj\"olund\inst{2}\orcidID{0000-0002-9099-3522} \and Peter Kimstrand\inst{2}\orcidID{0000-0001-9667-5595} \and 
Thomas~B.~Sch\"on\inst{1}\orcidID{0000-0001-5183-234X}}
\authorrunning{N. Gunnarsson et al.}
%
\institute{Department of Information Technology, Uppsala University, Sweden
\email{\{firstname\}.\{surname\}@it.uu.se}\\
 \and
 Elekta Instrument AB, Stockholm, Sweden \\ \email{\{firstname\}.\{surname\}@elekta.com}}

\maketitle

\begin{abstract}
Spatiotemporal imaging has applications in e.g. cardiac diagnostics, surgical guidance, and radiotherapy monitoring, In this paper, we explain the temporal motion by identifying the underlying dynamics, only based on the sequential images. Our dynamical model maps the inputs of observed high-dimensional sequential images to a low-dimensional latent space wherein a linear relationship between a hidden state process and the lower-dimensional representation of the inputs holds. For this, we use a conditional variational auto-encoder (CVAE) to nonlinearly map the higher-dimensional image to a lower-dimensional space, wherein we model the dynamics with a linear Gaussian state-space model (LG-SSM). The model, a modified version of the Kalman variational auto-encoder, is end-to-end trainable, and the weights, both in the CVAE and LG-SSM, are simultaneously updated by maximizing the evidence lower bound of the marginal likelihood. In contrast to the original model, we explain the motion with a spatial transformation from one image to another. This results in sharper reconstructions and the possibility of transferring auxiliary information, such as segmentation, through the image sequence. Our experiments, on cardiac ultrasound time series, show that the dynamic model outperforms traditional image registration in execution time, to a similar performance. Further, our model offers the possibility to impute and extrapolate for missing samples.
\end{abstract}

\section{Introduction}
Today, most medical imaging modalities support some form of time-resolved imaging. In some modalities, like ultrasound, it is the default mode of operation, while in others it is known under different names depending on the application, e.g. fluoroscopy~\cite{Wallace2008}, 4DCT~\cite{Pan2004} and 4D flow MRI \cite{Markl2012}. It is used for analysis, in e.g. cardiac diagnostics~\cite{angelini2001lv} and for guidance, monitoring and control in e.g. interventional and intraoperative percutaneous procedures and surgeries~\cite{blanco2005interventional,weil2007long} and image-guided radiation therapy~\cite{mittauer2018new,srinivasan2014applications}.

In this paper, we want to uncover the dynamics in a medical image time series based on nothing but the images themselves. Our aim is to learn a fast, nearly real-time, parameterized function, $h$, such that the image $y_t$ at time $t$ is described by the dynamical system
\begin{equation}
    y_t = h(y_{1:t-1}) + \epsilon_t,
    \label{eq:first}
\end{equation}
where $y_{1:t-1} = \{y_{1}, \dots y_{t-1}\}$ are the previously observed images and $\epsilon_t$ is noise. This representation makes it possible to predict $\hat{y}_t = h(y_{1:k})$ based on observations up to time $k$, i.e. to impute~($t<k$), filter~($t=k$), or extrapolate~($t>k$) images. Instead of reconstructing image $y_t$ directly from~(\ref{eq:first}) we model the motion of the dynamics, i.e. the spatial transformation $\varphi_t$ from a predefined reference image $y$. The reconstructed image is then acquired by transforming $y$ with the spatial transformation, $y_t = y \circ \varphi_t$. Modeling the image transformations instead of the images directly enables transferring auxiliary information, such as segmentation, from one image domain to another.

Prediction of 2D video sequences has lately shown impressive results~\cite{walker2021predicting,saxena2021clockworkvae,hafner2019dreamer} in computer vision. Video prediction methods are usually evaluated based on the visualization of the predicted frames. Here, and in the medical domain, we are interested in explaining the true spatial transformation that generates the dynamical motion.

Temporal medical images contain high-dimensional data where the dynamics, due to cyclic and deformable motion, is nonlinear. While the linear assumption is preferable for several reasons, including tractable filtering and smoothing posteriors, it is inappropriate for the raw image series. The idea of this paper is to reduce the high-dimensional data to a low-dimensional latent space wherein the linearity assumptions are valid. For this we combine techniques from generative and dynamic modeling and train the model end-to-end. In the literature, this is referred to as a deep state-space model (DSSM)~\cite{fraccaro2017disentangled,karl2016deep,krishnan2017structured}.

In the medical domain, there are many examples of motion modeling based on motion proxies \cite{mcclelland2013respiratory,verma2010survey} such as tracking implanted markers using X-ray imaging or surface tracking. Those methods are limited to the accuracy and applicability of the motion proxy. To the best of our knowledge, the only example where the motion model is based directly and exclusively on the temporal medical images under study is the recent work by \cite{krebs2021learning}. They model the dynamics in a low-dimensional probabilistic space using a temporal convolutional network~\cite{bai2018empirical} and a Gaussian process prior. In their model, the prediction length is restricted based on the motion matrix. We overcome this by modeling the dynamics as a first-order Markov process.
\section{Background}
The Kalman variational auto-encoder~(Kalman VAE)~\cite{fraccaro2017disentangled} is an unsupervised model for high-dimensional sequential data that, most likely, undergoes nonlinear dynamics. Higher-dimensional observed images, $\vy = \{y_t\}_{t=1}^T$ are non-linearly embedded into a lower-dimensional space using a variational auto-encoder. The dynamics of the lower-dimensional features, $\vx =\{x_t\}_{t=1}^T$ is modeled with a linear Gaussian state-space model based on a state-space process $\vz =\{z_t\}_{t=1}^T$. Below we explain the variational auto-encoder and the linear Gaussian state-space model, and then how those are combined into the framework of Kalman VAE.

\subsection{Variational auto-encoder}
Similar to traditional auto-encoders~\cite{kramer1991nonlinear}, variational auto-encoders~(VAEs) embed the input $y$ in a lower-dimensional latent space $x$ using an encoder, $E_\phi$ and reconstruct the original input with a decoder, $D_\theta$. They differ in that VAEs are generative and reconstruct the data distribution $p_\theta(y)$ instead of a single sample $y$. In this case the true posterior $p_\theta(x \mid y)$ is intractable. By approximating the variational posterior as a multivariate Gaussian, $q_\phi (x \mid y)$ $= \mathcal{N} (x \mid \mu^{\text{enc}}, \Sigma^{\text{enc}})$, where $\mu^{\text{enc}}$ and $\Sigma^{\text{enc}}$ are the outputs from the encoder, it is possible to sample from the variational approximation. From the KL divergence between the approximate and the true posterior we can obtain a lower bound on the true likelihood 
\begin{align}
    \begin{split}
    \log p_\theta(y) \geq \mathbb{E}_{q_\phi(x \mid y)} \Big[ \log p_\theta(y \mid x) + \log \dfrac{p_\theta(x)}{q_\phi(x \mid y)} \Big],
    \end{split}
    \label{eq:loss_vae_paper}
\end{align}
where the prior over the latent space is usually chosen to be a multivariate Gaussian $p_\theta(x) = \gN(x \mid 0, \text{I})$. This lower bound is called the evidence lower bound (ELBO) and can be estimated by sampling~\cite{Kingma2014}.

\subsection{Linear Gaussian state-space model}
In a VAE, each sample, $x_t\in \R^L$, from the approximate posterior is normally distributed with mean, $\mu_t^{\text{enc}}$, and covariance $\Sigma_t^{\text{enc}}$.  
If we assume Gaussian noise, it follows that the state-space vector, $z_t$, in our linear state-space model also follows a normal distribution. More precisely, we have a linear Gaussian state-space model (LG-SSM),
\begin{align}
\begin{split}
    p(z_t \mid z_{t-1}) & = \mathcal{N}(z_t\mid A z_{t-1}, Q), \\
    p(x_t \mid z_t) & = \mathcal{N}(x_t\mid C z_t, R),
\end{split}
\label{eq:lgssm}
\end{align}
where $Q$ and $R$ are covariance matrices for the process and measurement noise, respectively. Given an initial guess $z_1 \sim \mathcal{N} (z_1 \mid \mu_{1}, P_{1})$, the joint probability distribution can be expressed using the LG-SSM model (from~(\ref{eq:lgssm})),
\begin{align}
\begin{split}
    p(\vx, \vz) & = p(\vx \mid \vz) p(\vz)  = p(z_1) \prod_{t=1}^T  p(x_t \mid z_t) \prod_{t=2}^T p(z_t \mid z_{t-1}). \label{eq:joint_lgssm}
\end{split}
\end{align}
Given observations $\vx$ the mean and covariance of the state-space variables is analytically tractable using a Kalman filter~\cite{kalman1960new}, $\mu_{t \mid t}, P_{t \mid t}$, and a Rauch-Tung-Striebel (RTS) smoother~\cite{rauch1965maximum}, $\mu_{t \mid T}, P_{t \mid T}$. 

\subsection{Kalman variational auto-encoder}
In the Kalman VAE a VAE is used to reduce the dimension of the image time series distribution wherein the dynamics are represented linearly using an LG-SSM in the latent space. The goal is to describe the dynamics of the system in the latent space with an LG-SSM and use the decoder to reconstruct the image time series. 

Similar to a regular VAE, the ELBO can be derived from the KL divergence between the approximate and true posterior. In a Kalman VAE the approximate posterior is given by
\begin{align}
    q_{\phi, \gamma}(\vx,\vz \mid \vy) = p_\gamma(\vz \mid \vx) q_\phi(\vx \mid \vy),
\end{align}
and the true posterior is proportional to
\begin{align}
     p_{\theta, \gamma}(\vx,\vz,\vy) \propto p_\theta(\vy \mid \vx)p_\gamma(\vx,\vz).
\end{align}
For a single time series $\vy = \{y_t\}_{t=1}^T$ the ELBO is given by, $\gL^{\text{KVAE}}_{\phi, \theta, \gamma}$
\begin{equation}
    \begin{split}
        \gL^{\text{KVAE}}_{\phi, \theta, \gamma} = 
        \E_{q_{\phi}(\vx \mid \vy)}
        \Big[
        \log \dfrac{
        p_\theta(\vy\mid \vx)}
        {q_\phi(\vx\mid \vy)} +  \E_{p_\gamma(\vz\mid \vx)}\Big[ \log \dfrac{p_\gamma(\vx, \vz)}{p_\gamma(\vz \mid \vx)}\Big]\Big],
    \end{split}
    \label{eq:elbo_kvae_paper}
\end{equation}
where $\phi$ and $\theta$ are the encoder and decoder parameters, respectively, and $\gamma = \{$ $A$, $C$, $R$, $Q$, $\mu_{1}, P_{1}\}$ are the LG-SSM parameters. It is possible to sample $(\Tilde{\mathbf{x}}, \Tilde{\mathbf{z}})$ by first sampling $\Tilde{\mathbf{x}} \sim q_\theta(\mathbf{x} \mid \mathbf{y})$ and then conditionally sampling $\Tilde{\mathbf{z}} \sim p_\gamma(\mathbf{z} \mid \Tilde{\mathbf{x}})$. Notice that $p_\gamma(\mathbf{z} \mid \Tilde{\mathbf{x}})$ is tractable using the Kalman smoother algorithm and the joint distribution $p_\gamma(\mathbf{x}, \mathbf{z})$ is given by~(\ref{eq:joint_lgssm}). With the re-parameterization trick~\cite{Kingma2014} the model can be trained end-to-end to minimize the negative ELBO using e.g. stochastic gradient descent.
\section{Method}
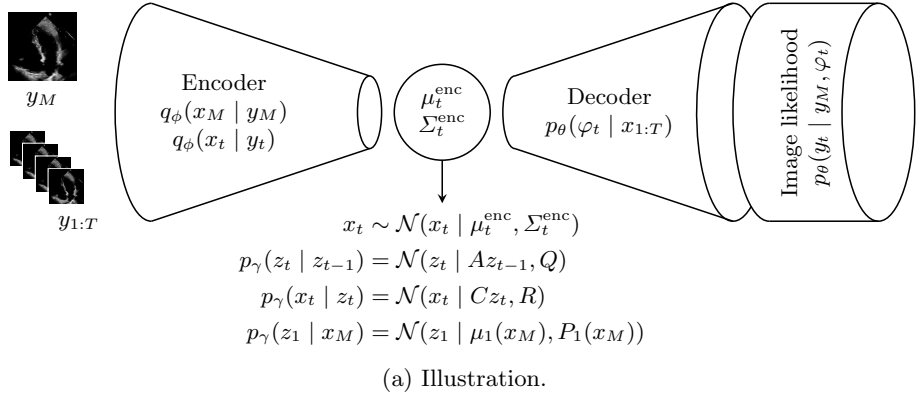
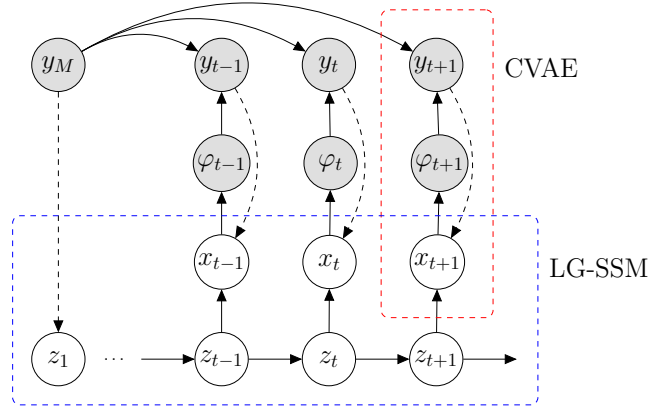
\begin{figure}[htp!]
\centering
\subfloat[Illustration.\label{ill}]{%
\resizebox{1.0\linewidth}{!}{\subfile{../tikz/model}}}
\\
\centering
\subfloat[Graphical representation: The generative and inference model is represented with black solid and dashed lines, respectively.\label{graphical}]{%
\resizebox{0.7\linewidth}{!}{\begin{tikzpicture}[x=1.0cm,y=0.8cm]
    \node[obs, minimum size=1cm] (y1) {\Large $y_{t-1}$};
    \node[obs, minimum size=1cm, right=of y1] (y2) {\Large $y_{t}$};
    \node[obs, minimum size=1cm, right=of y2] (y3) {\Large $y_{t+1}$};
    
    \node[const, left=of y1]  (t1_0) {}; %
    \node[obs, minimum size=1cm, left=of t1_0] (y0) {\Large $y_M$};
    
    \node[obs, minimum size=1cm, below=of y1] (phi1) {\Large $\varphi_{t-1}$};
    \node[obs, minimum size=1cm, right=of phi1] (phi2) {\Large $\varphi_{t}$};
    \node[obs, minimum size=1cm, right=of phi2] (phi3) {\Large $\varphi_{t+1}$};
    
    \node[latent, minimum size=1cm, below= of phi1] (x1) {\Large $x_{t-1}$};
    
    \node[latent, minimum size=1cm, right=of x1] (x2) {\Large $x_{t}$};
    \node[latent, minimum size=1cm, right=of x2] (x3) {\Large $x_{t+1}$};

    \node[latent, minimum size=1cm, below= of x1] (z1) {\Large $z_{t-1}$};
    \node[latent, minimum size=1cm, right=of z1] (z2) {\Large $z_{t}$};
    \node[latent, minimum size=1cm, right=of z2] (z3) {\Large $z_{t+1}$};
    \node[const, left=of z1]  (t_0) {}; %
    \node[const, right=of z3]  (t_N) {}; %
    \node[latent, minimum size=1cm, left= of t_0] (z0) {\Large $z_1$};
    
    \path (x1) edge [-triangle 45] node {} (phi1) ;
    \path (phi1) edge[-triangle 45] node {} (y1);
    \path (x2) edge [-triangle 45] node {} (phi2) ;
    \path (phi2) edge [-triangle 45] node {} (y2);
    \path (x3) edge [-triangle 45] node {} (phi3) ;
    \path (phi3) edge [-triangle 45] node {} (y3);

    \path (z1) edge [-triangle 45] node {} (x1) ;
    \path (z2) edge [-triangle 45] node {} (x2) ;
    \path (z3) edge [-triangle 45] node {} (x3) ;
    
    \path (t_0) edge [-triangle 45] node {} (z1) ;
    \path (z1) edge [-triangle 45] node {} (z2) ;
    \path (z2) edge [-triangle 45] node {} (z3) ;
    \path (z3) edge [-triangle 45] node {} (t_N);
    
    \path (y0) edge [bend left, -triangle 45] node {} (y1);
    \path (y0) edge [bend left, -triangle 45] node {} (y2);
    \path (y0) edge [bend left, -triangle 45] node {} (y3);
    
    \path (y1) edge [bend left, dashed, -triangle 45] node {} (x1);
    \path (y2) edge [bend left, dashed, -triangle 45] node {} (x2);
    \path (y3) edge [bend left, dashed, -triangle 45] node {} (x3);
    
    \path (y0) edge [dashed, -triangle 45] node {} (z0);
    
    \node (b) at ($(z0)!0.35!(z1)$) {$\dots$};

    \gate[red, rounded corners,inner sep=15pt] {vae} {(y3)(x3)} {} ; %
    \gate[blue, rounded corners,inner sep=10pt] {ssm} {(z0)(t_0)(x1)(x2)(x3)(t_N)(z1)(z2)(z3)} {} ; %
    
    \node[const, right=of vae, xshift=-0.8cm, yshift=1.8cm]  () {\Large CVAE};
    \node[const, right=of ssm, xshift=-0.8cm, yshift=0.8cm]  () {\Large LG-SSM};
\end{tikzpicture}}}
\caption{Illustration~\protect\subref{ill} and graphical representation~\protect\subref{graphical} of our model. Similar to a Kalman VAE, we learn a lower-dimensional linear-Gaussian state-space model from data, but unlike a Kalman VAE our model operates on image transformations $\varphi_t$ and not images.}
\label{fig:our_model}
\end{figure}

Unlike a standard Kalman VAE, our method operates in image \emph{transformations} and not image per se. More specifically, we assume that the images $\vy$ are observations from the dynamical state-space model
\begin{align} \label{eq:dyn}
    \begin{split}
    \varphi_{t} & = f(\varphi_{t-1}) + \epsilon_t, \\
    y_t & = g(\varphi_t, y_{M}) + \eta_t, 
    \end{split}
\end{align}
where $y_{M} \in \gM$ is a reference image, $\varphi_t$ is the image transformation from the stationary image domain to the image domain at time $t$ and $\epsilon_t$, $\eta_t$ are process and measurement noise, respectively. We hereby describe the model as a sequence of spatiotemporal image registrations. To be consistent with the medical image community we define $y_M$ as the moving image. An illustration and graphical representation of our model are provided in Figure~\ref{fig:our_model}.


The dynamics of our model are driven by the transformation $\varphi_t$ with respect to the spatial information in $y_M$. To include the spatial information of $y_M$ we have used a modified Kalman VAE, where the mappings to and from the latent representation are given by a conditional variational auto-encoder (CVAE) \cite{kingma2014semi}, conditioned on the moving image $y_M$. The motion model in the latent space is modeled with an LG-SSM. Spatial information is included in the motion model by condition the initial state prior on the latent representation, $x_M$ of the moving image. The initial prior in the LG-SSM is then given by $p_\gamma(z_1 \mid x_M) = \gN(z_1 \mid \mu_1(x_M),P_1(x_M))$. For this, we use a fully connected neural network to estimate the mean and variance of this distribution. The parameters of our model can be jointly updated by maximizing the ELBO, $\gL_{\phi, \theta, \gamma}$ 
\begin{equation}
    \begin{split}
        \E_{q_{\phi}(\vx, x_M \mid \vy, y_M)}\Big[ 
        \log \dfrac
        {p_\theta(\vy\mid \vx, y_M)}
        {q_\phi(\vx, x_M\mid \vy, y_M)}  + 
         \E_{p_\gamma(\vz\mid \vx, x_M)}\Big[ \log \dfrac{p_\gamma(\vx, \vz \mid x_M)}{p_\gamma(\vz \mid \vx, x_M)}\Big]\Big],
    \end{split}
    \label{eq:elbo}
\end{equation}
where the ELBO contains of one VAE-part and one LG-SSM-part. For the VAE-part, the generative model is given by $p_\theta(\vy \mid \vx, x_M) = \prod_{t=1}^T p_\theta(y_t \mid x_t, y_M)$ and the posterior $q_\phi(\vx, x_M \mid \vy, y_M) = q_\phi(x_M \mid y_M)\prod_{t=1}^T q_\phi(x_t \mid y_t)$. In the LG-SSM part the prior is given by 
\begin{equation}
 p_\gamma(\vx, \vz \mid x_M)  = p_\gamma(z_1 \mid x_M) \prod_{t=1}^T p_\gamma(x_t \mid z_t)  \prod_{t=2}^T p_\gamma(z_t \mid z_{t-1})   
\end{equation}
and the exact conditional posterior $p_\gamma(\vz \mid \vx,  x_M) = \prod_{t=1}^T p_\gamma(z_t \mid \vx, x_M)$ can be obtained with RTS smoothing.

\subsection{Likelihood}
In the Kalman VAE the likelihood is assumed to come from some parametric family of distributions, parameterized by the decoder network $D_\theta(x_t)$. Instead of modeling the likelihood of the image intensity we model the motion, i.e. the likelihood of the image transformation $\varphi_t$. From the decoder this is modeled 
\begin{align}
    p_\theta(\varphi_t \mid x_t) & = \gN(\varphi_t \mid \mu_t^{\text{dec}}, \Sigma_t^{\text{dec}}),
\end{align}
as a multivariate Gaussian distribution, i.e. the decoder outputs $\mu_t^{\text{dec}}$ and $\Sigma_t^{\text{dec}}$. A common disadvantage of image-based VAEs is blurry reconstructions due to the restrictive assumption of a Gaussian likelihood~\cite{dosovitskiy2016generating}. High frequencies, like sharp edges and fine details in the images are often missed in the reconstructions~\cite{cai2019multi}. In the space of image transformations, on the other hand, Gaussian assumptions are quite reasonable. Existing methods use Gaussian kernels to regularize the displacement field and avoid improbable transformations~\cite{thirion1998image}. We then estimate the image likelihood as a noisy observation of the transformed image, $y_M \circ \varphi_t$, i.e. 
\begin{align}
    p_\theta(y_t \mid y_M, \varphi_t) & = \gN(y_t \mid y_M \circ \varphi_t, \sigma_t^2 I),
\end{align}
for some noise $\sigma_t$.  

\section{Experiments}
\subsection{Dataset}
In our experiments, we used the EchoNet-Dynamic dataset~\cite{ouyang2020video}. The dataset consists of $10 030$ 2D ultrasound echocardiogram time series with $112 \times 112$ pixel frames. We fixed the time horizon to $50$ time steps ($T=50$), which corresponded to 1 second and approximately included one cardiac cycle. We extracted one sequence per time series with a randomized start position. Sequences shorter than the fixed horizon were removed ($315$). We allocated $7 220$ sequences for training, $1 237$ for testing and $1 258$ for validation. The image intensity was normalized to $[0,1]$ for all images.   


\subsection{Metrics}
We evaluate our model using three metrics, the Dice coefficient~\cite{dice1945measures}, the percentage of non-positive Jacobians (Jacobian determinants), and the execution time.

The data sequence included human expert segmentation of the left ventricle at two different time points. We used the image for the first human expert segmentation as moving image $y_M$, and estimated the transformation for the sequence, including the time step for the other human expert segmented image. We can then measure the quality of transformation with the Dice coefficient, the overlap between the human expert segmentation, $S_t$ and our estimation, $S_M \circ \varphi_t$,
\begin{align}
    \text{Dice} = 2\dfrac{|S_t \cap(S_M \circ \varphi_t)|}{|S_t| + |S_M \circ \varphi_t|},
\end{align}
where a perfect overlap is indicated with a Dice coefficent of $1$, and $0$ for no overlap.

The percentage of non-positive elements in the Jacobian is an indicator of how topology-preserving the transformation is, e.g. if $|J_\varphi(p)| > 0$ for all points $p$ the transformation $\varphi$ is diffeomorphic. The Jacobian, $|J_\varphi|$ is defined by
\begin{align}
    \renewcommand\arraystretch{2}
    |J_\varphi| = \begin{vmatrix}
\dfrac{\partial \varphi_x}{\partial x} & \dfrac{\partial \varphi_x}{\partial y}\\
\dfrac{\partial \varphi_y}{\partial x} & \dfrac{\partial \varphi_y}{\partial y}
\end{vmatrix}.
\end{align}

Although the execution time is hardware dependent, we believe it is a relevant metric since we aim for real-time estimates.


\subsection{Implementation Details}
\subsubsection*{Encoder}
The encoder consists of downsampling convolutional levels with $16$, $32$, $64$ and $128$ filters respectively, where the feature maps at each level are first downsampled using a 2-stride convolutional layer followed by a batch normalization~\cite{ioffe2015batch} and a leaky ReLU~\cite{maas2013rectifier} activation function. Furthermore, we use two 1-stride convolutional layers with residual connections~\cite{he2016deep} before the feature map is downsampled to the next level. The output of the last level is flattened and two dense layers are used---one for the mean, $\{\mu_t^{\text{enc}}\}_{t = 1}^T$ and one for the variances, $\{\left({\sigma_t^\text{enc}}\right)^2\}_{t=1}^T$. 
\subsubsection*{LG-SSM}
In the LG-SSM we use a dimension of $16$ for our observations, $x_t \in \mathbb{R}^{16}$, and $32$ for the state space, $z_t \in \mathbb{R}^{32}$. The mean and variance of the initial prior was estimated with a 3 layer fully connected neural network with $16, 16$ and $32$ units.
\subsubsection*{Decoder} The decoder layers mirror the encoder with the same filter sizes and residual connections at each level. We let the decoder estimate $\mu_t^{\text{dec}}$, the mean of the transformation likelihood, and we used a fixed covariance $\Sigma_t^{\text{dec}} = 0.01^2 I$. When sampling from the transformation likelihood we apply a mask which corresponds to the cone-shaped field-of-view of the ultrasound image. 

The mean of the image likelihood is estimated by the sample from the transformation likelihood and applying the transformation on a moving image $y_M$. For this, we use a spatial transformation module~\cite{jaderberg2015spatial,balakrishnan2019tmi}. In this example, we use the first image in the time sequence as a moving image. The variance for the image likelihood was fixed to $\sigma_t^2 = 0.01^2$.
\subsubsection*{Training}
The model is trained end-to-end using importance sampling to maximize the ELBO in (\ref{eq:elbo}) by jointly updating all parameters $\{\theta, \phi, \gamma\}$ in the model. We use a monotonic annealing schedule weight~\cite{fu2019cyclical} to the posteriors in the loss function. As an optimizer, we use Adam with exponential decay with factor $0.85$ every $20$ epochs and an initial learning rate of $10^{-4}$. We use a batch size of $4$ image time series and train the model for $50$ epochs on a single Nvidia GeForce GTX 1080 Ti graphic card ($\approx 17$ hours training). We have implemented our method into the Tensorflow framework~\cite{tensorflow2015_whitepaper} and the code is publicly available
\footnote{
\url{https://github.com/ngunnar/med-dyn-reg}
}.
\subsubsection*{Evaluation}
When evaluating the result we use samples from the filtered distribution $x_t \sim p_\gamma(x_t \mid z_{1:t})$.
We found that our initial model produced transformations where the Jacobian was frequently negative. We therefore apply regularization to the displacement field using a Gaussian kernel, $g_\sigma$
\begin{align}
    \hat{\varphi_t} = \varphi_t \star g_\sigma,
\end{align}
as an alternative version of our model.

\subsection{Result}
We compare our model with the Demons algorithm~\cite{thirion1998image}. Demons is a popular and fast iterative method for image registration. Since it is iterative, the number of iterations is manually selected. We choose the number of iterations to 40, which gives a fair trade-off between accuracy and computational efficiency in this experiment. We also compare the result with no applied registration to verify the improvements. The average result on the validation data is shown in Table~\ref{tab:result}.   

\begin{table*}[ht!]
    \centering
    \resizebox{\textwidth}{!}{\begin{tabular}{@{}lccccc@{}}
    \toprule
    \textbf{Model} & \textbf{info} & \textbf{Avg. Dice $\uparrow$} & \% $\mathbf{|J_\varphi|\leq 0 \downarrow}$ & \textbf{GPU (s) $\downarrow$} & \textbf{CPU (s) $\downarrow$}\\
    \toprule
        our & filtered & $0.81\pm0.06$ & $2.7\% \pm 2.7\%$ & $0.09 \pm 0.3\cdot 10^{-2}$  & $0.22 \pm 0.5\cdot 10^{-2}$\\
        our reg & filtered  & $0.81\pm0.06$ & $0.5\% \pm 1.0\%$ & $0.09 \pm 0.3\cdot 10^{-2}$ & $ 0.33 \pm 1 \cdot 10^{-2} $  \\
        Demons & 40 iter.  & $0.81 \pm 0.07$ & $1.2\% \pm 1.0\%$ & - & $1.64\pm 2 \cdot 10^{-2}$ \\
        None  & - &$0.74\pm0.07$ & - & - & -\\
    \bottomrule
    \end{tabular}}
    \caption{We compare the average Dice, percentage of negative Jacobian determinants, and execution time with the Demons algorithm. The time is the average execution time for the entire sequence.}
    \label{tab:result}
\end{table*}

\subsubsection{Visualization}
\begin{figure*}[ht!]
    \centering
    \resizebox{1.0\linewidth}{!}{
\def\b{0.11}
\resizebox{\textwidth}{!}{
\begin{tabular*}{\textwidth}{cccccccccccc}%
    \toprule
     & & $t=1$ & $t=6$ & $t=11$ & $t=16$ & $t=21$ & $t=26$ & $t=31$ & $t=36$\\
    \toprule
    \rotatebox[origin=c]{90}{\small \textbf{Data}} & & 
    \raisebox{-0.5\height}{\includegraphics[width=\b\textwidth]{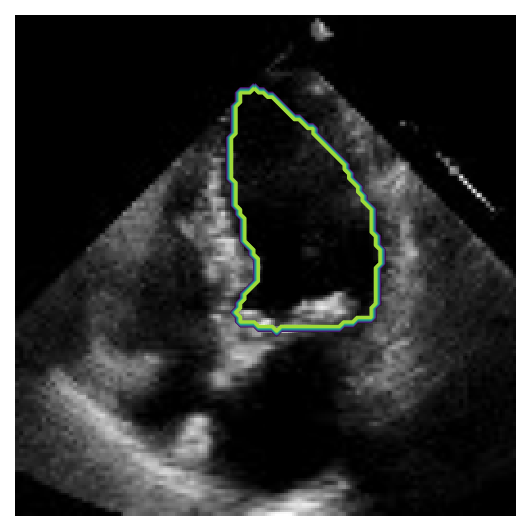}} & 
    \raisebox{-0.5\height}{\includegraphics[width=\b\textwidth]{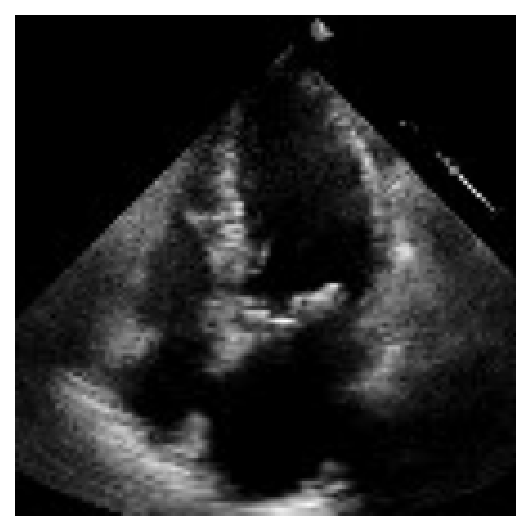}} & 
    \raisebox{-0.5\height}{\includegraphics[width=\b\textwidth]{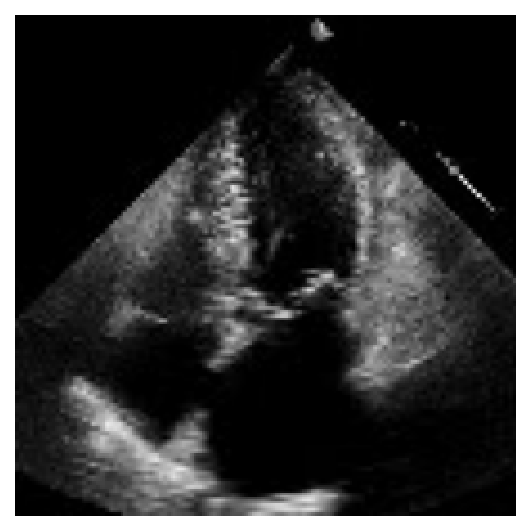}} &  
    \raisebox{-0.5\height}{\includegraphics[width=\b\textwidth]{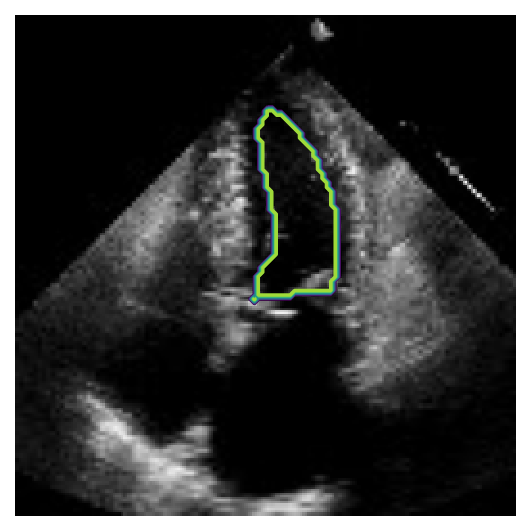}} &
    \raisebox{-0.5\height}{\includegraphics[width=\b\textwidth]{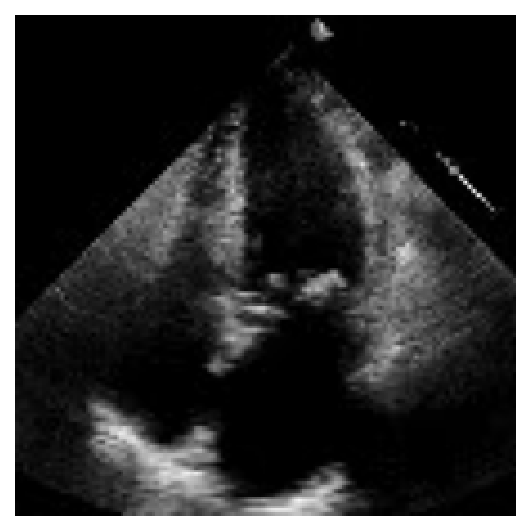}} & 
    \raisebox{-0.5\height}{\includegraphics[width=\b\textwidth]{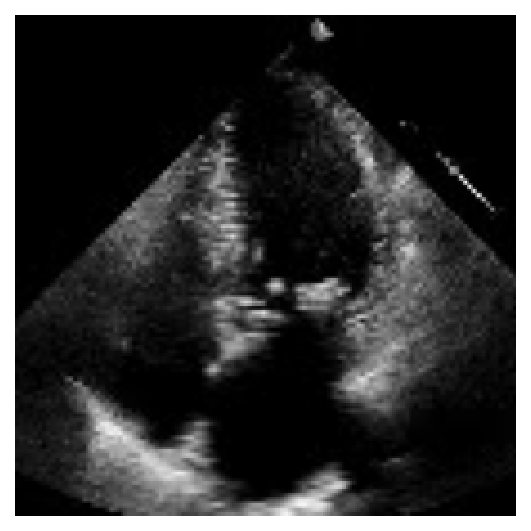}} &  
    \raisebox{-0.5\height}{\includegraphics[width=\b\textwidth]{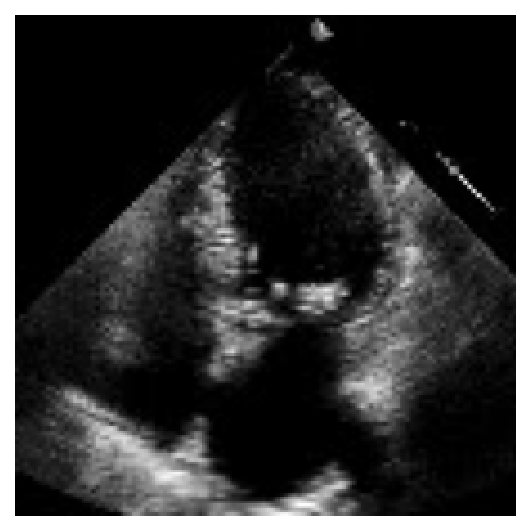}} &  
    \raisebox{-0.5\height}{\includegraphics[width=\b\textwidth]{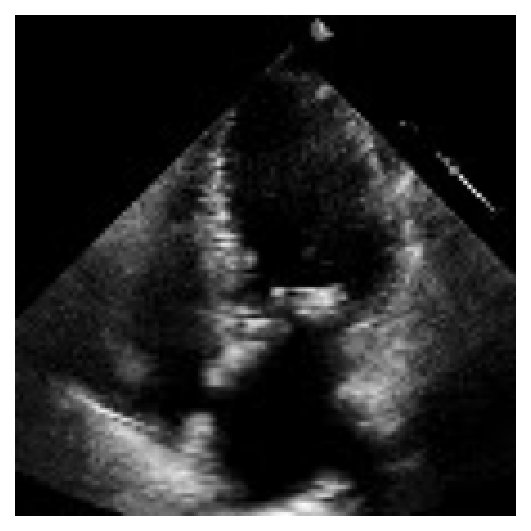}} 
    \\
    \midrule
    \multirowcell{2}{\rotatebox[origin=c]{90}{\small \textbf{Reconstruction}}} &
    \rotatebox[origin=c]{90}{\scriptsize Our} &
    \raisebox{-0.5\height}{\includegraphics[width=\b\textwidth]{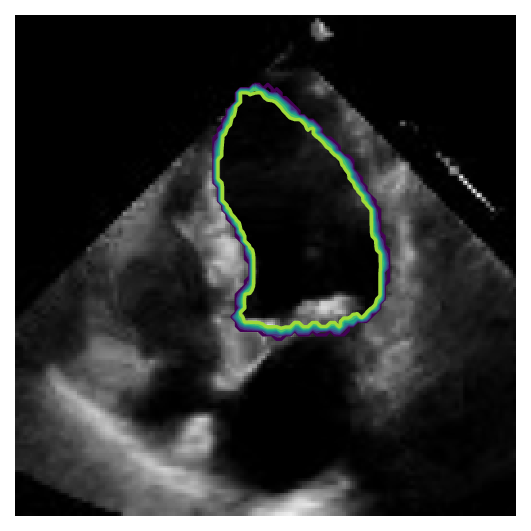}} & 
    \raisebox{-0.5\height}{\includegraphics[width=\b\textwidth]{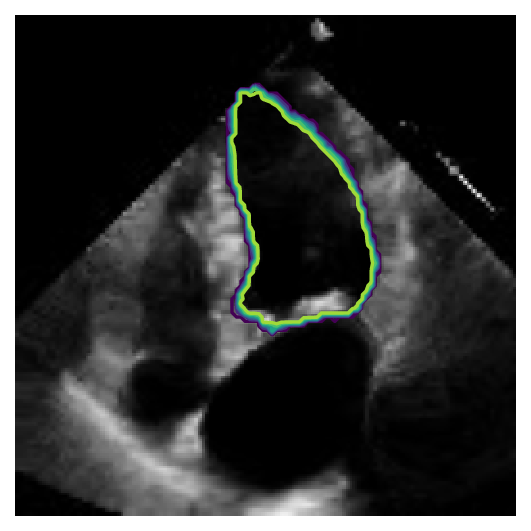}} & 
    \raisebox{-0.5\height}{\includegraphics[width=\b\textwidth]{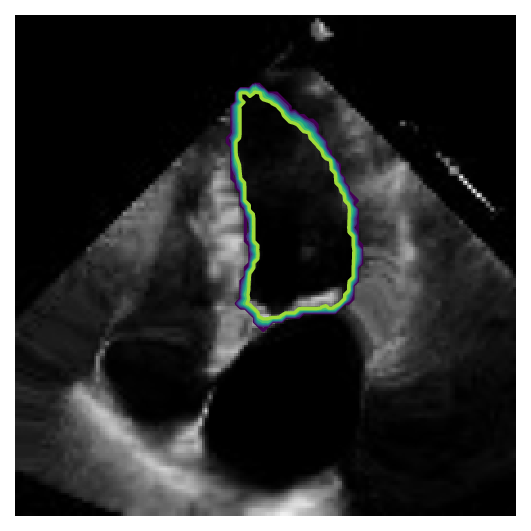}} &  
    \raisebox{-0.5\height}{\includegraphics[width=\b\textwidth]{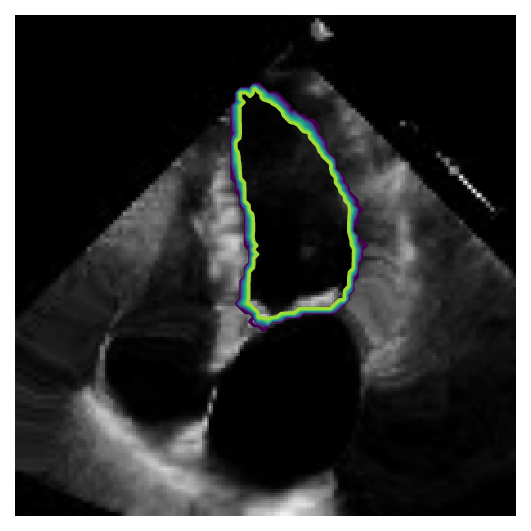}} &
    \raisebox{-0.5\height}{\includegraphics[width=\b\textwidth]{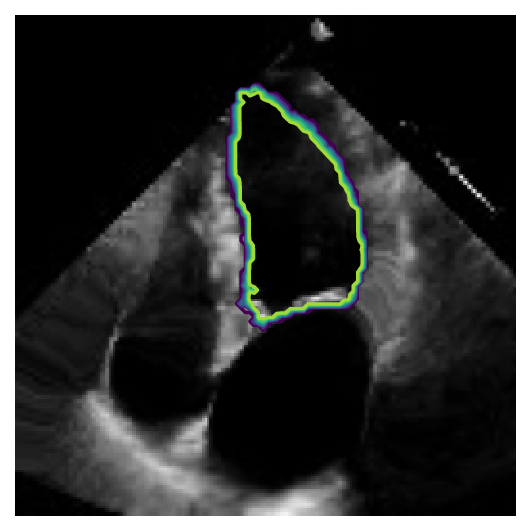}} & 
    \raisebox{-0.5\height}{\includegraphics[width=\b\textwidth]{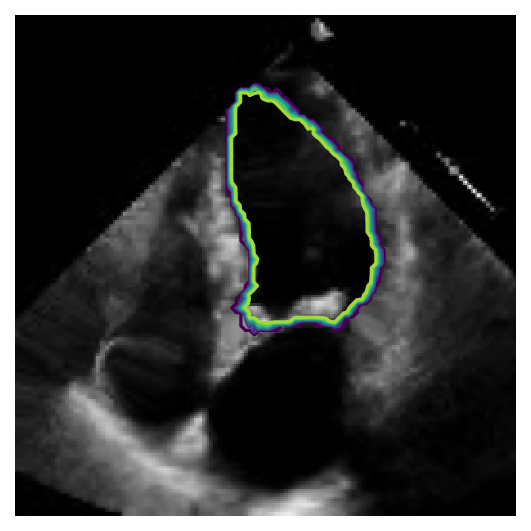}} & 
    \raisebox{-0.5\height}{\includegraphics[width=\b\textwidth]{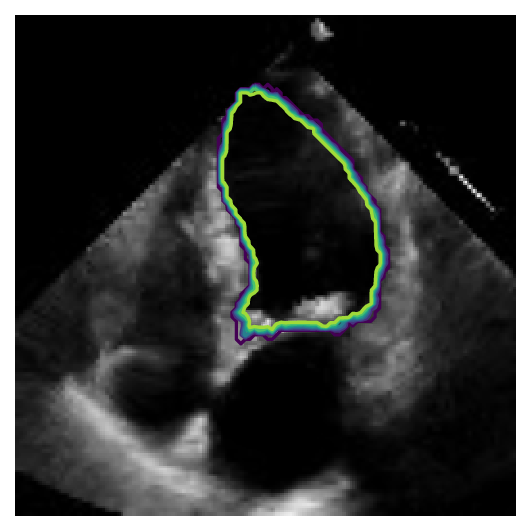}} &
    \raisebox{-0.5\height}{\includegraphics[width=\b\textwidth]{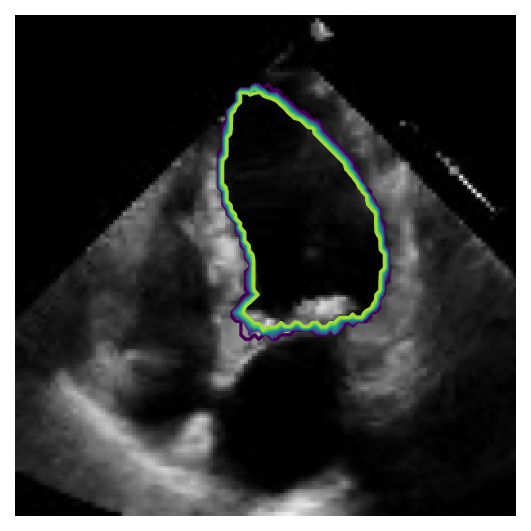}} 
    \\
    & 
    \rotatebox[origin=c]{90}{\scriptsize Demons} & 
    \raisebox{-0.5\height}{\includegraphics[width=\b\textwidth]{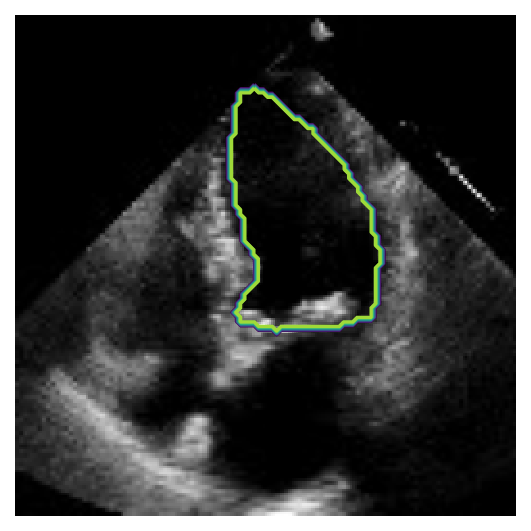}} &
    \raisebox{-0.5\height}{\includegraphics[width=\b\textwidth]{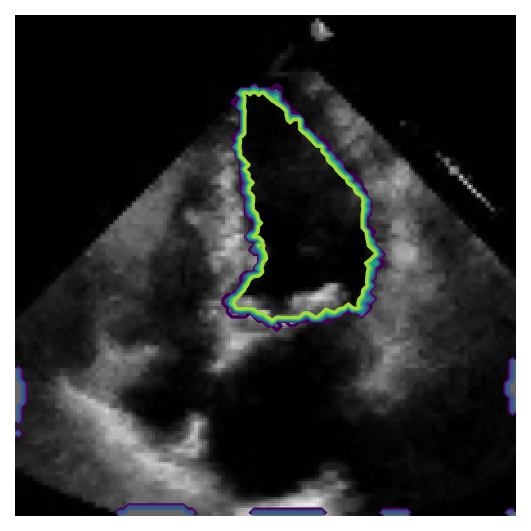}} &
    \raisebox{-0.5\height}{\includegraphics[width=\b\textwidth]{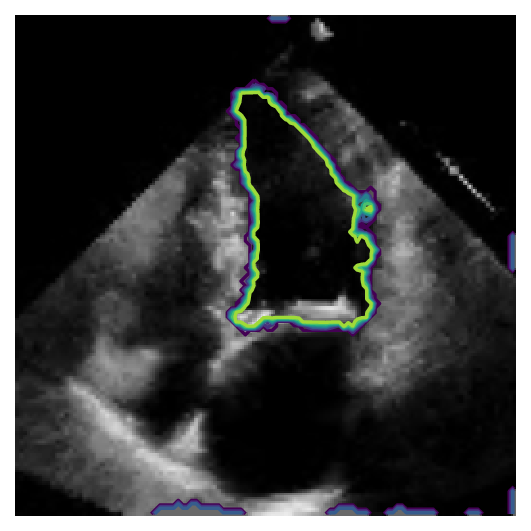}} &
    \raisebox{-0.5\height}{\includegraphics[width=\b\textwidth]{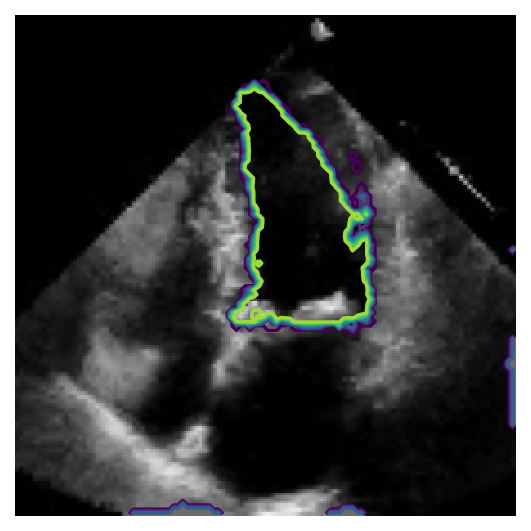}} &
    \raisebox{-0.5\height}{\includegraphics[width=\b\textwidth]{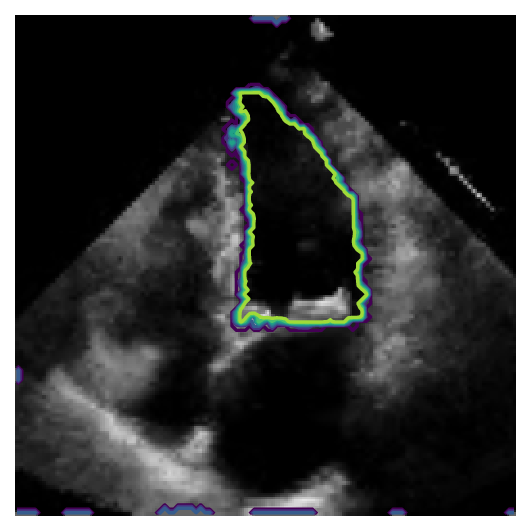}} &
    \raisebox{-0.5\height}{\includegraphics[width=\b\textwidth]{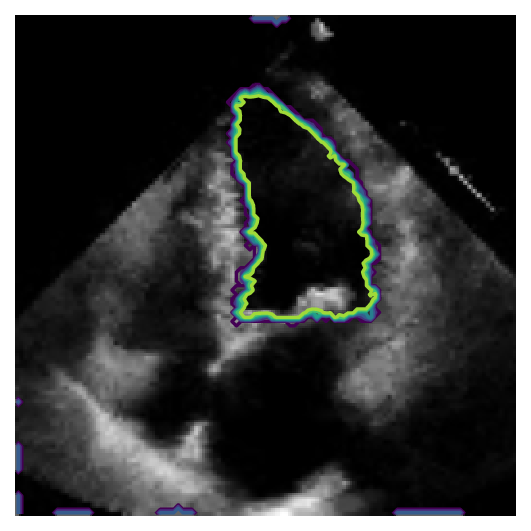}} &
    \raisebox{-0.5\height}{\includegraphics[width=\b\textwidth]{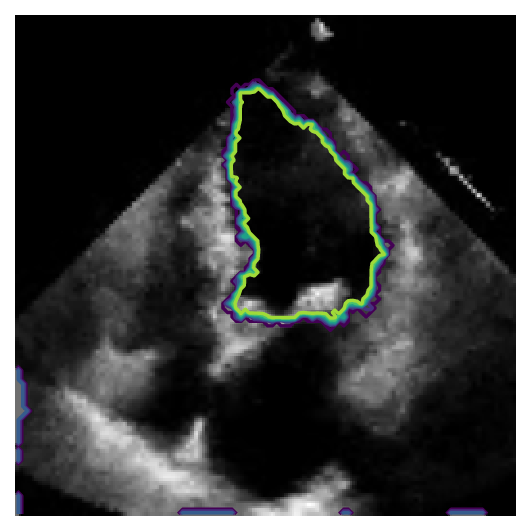}} &
    \raisebox{-0.5\height}{\includegraphics[width=\b\textwidth]{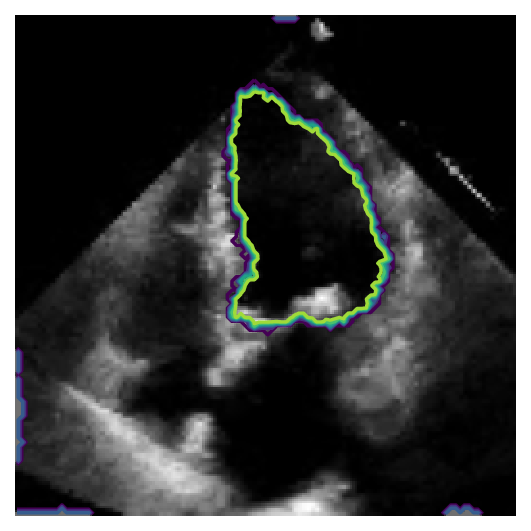}}
    \\ \\
    \midrule
    \multirowcell{4}{\rotatebox[origin=c]{90}{\small \textbf{Jacobian}}} &
    \rotatebox[origin=c]{90}{\scriptsize Our} &   
    \raisebox{-0.5\height}{\includegraphics[width=\b\textwidth]{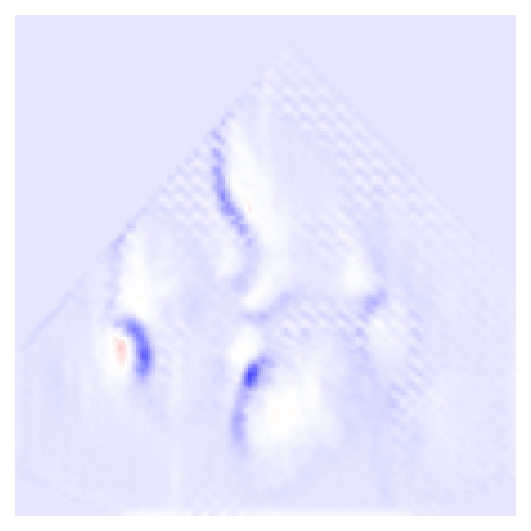}} & 
    \raisebox{-0.5\height}{\includegraphics[width=\b\textwidth]{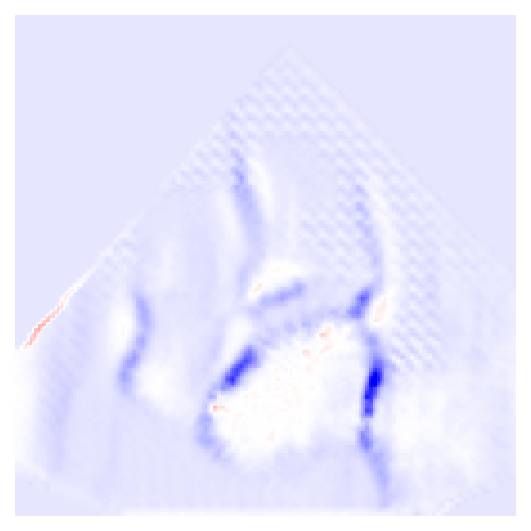}} & 
    \raisebox{-0.5\height}{\includegraphics[width=\b\textwidth]{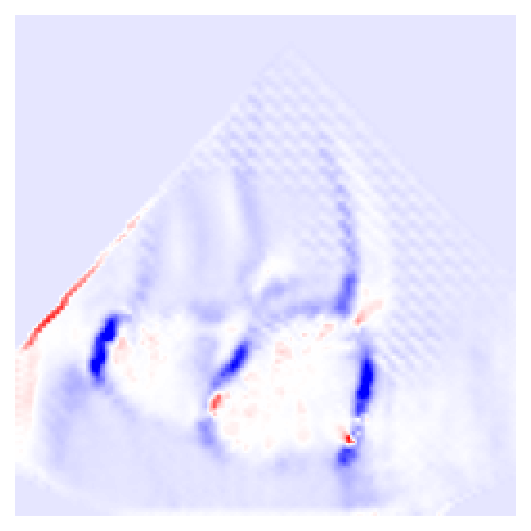}} &  
    \raisebox{-0.5\height}{\includegraphics[width=\b\textwidth]{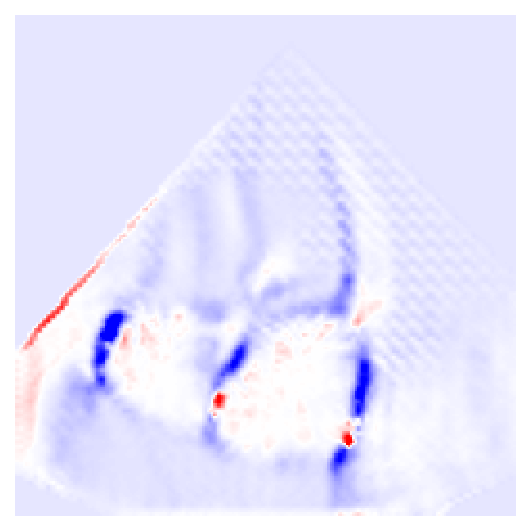}} &
    \raisebox{-0.5\height}{\includegraphics[width=\b\textwidth]{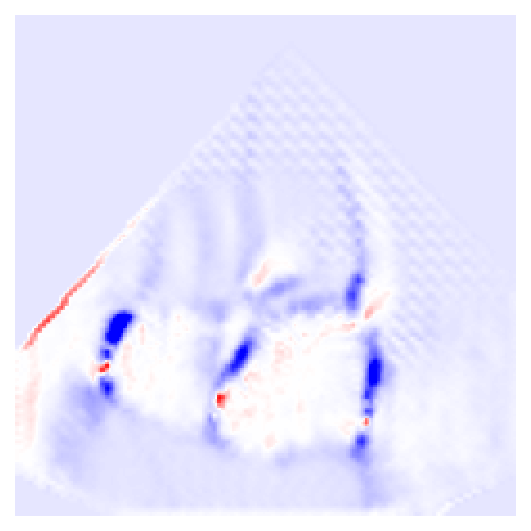}} & 
    \raisebox{-0.5\height}{\includegraphics[width=\b\textwidth]{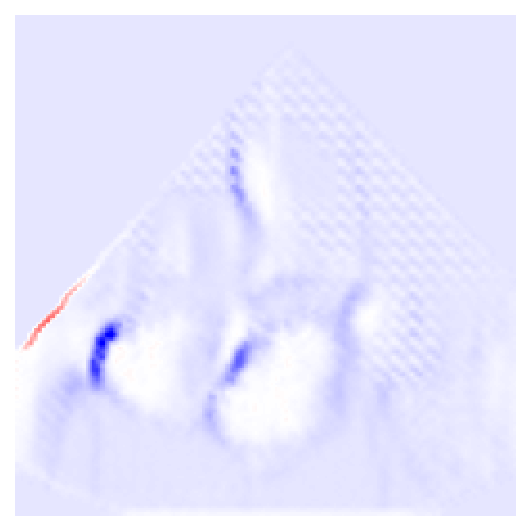}} & 
    \raisebox{-0.5\height}{\includegraphics[width=\b\textwidth]{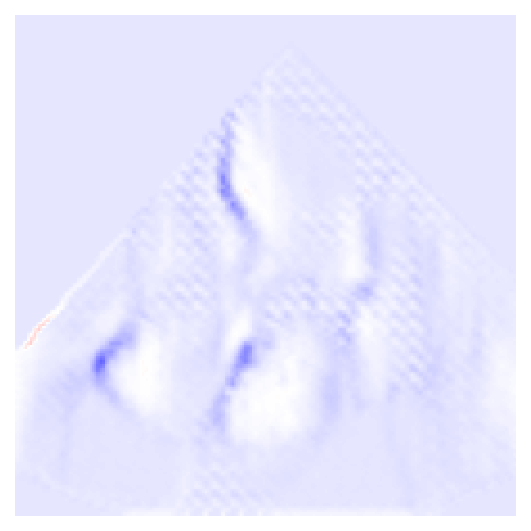}} &  
    \raisebox{-0.5\height}{\includegraphics[width=\b\textwidth]{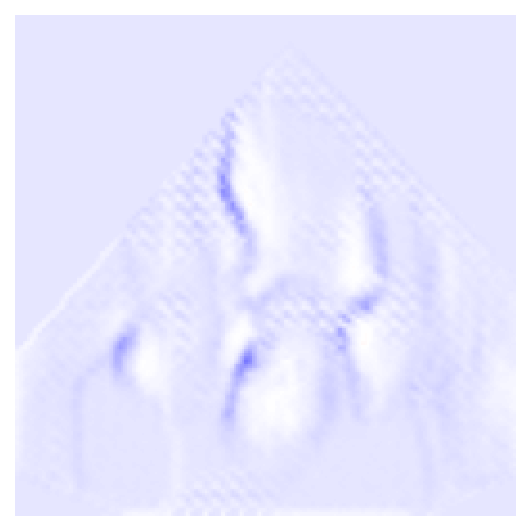}} 
    \\
    & 
    \rotatebox[origin=c]{90}{\scriptsize Demons} &
    \raisebox{-0.5\height}{\includegraphics[width=\b\textwidth]{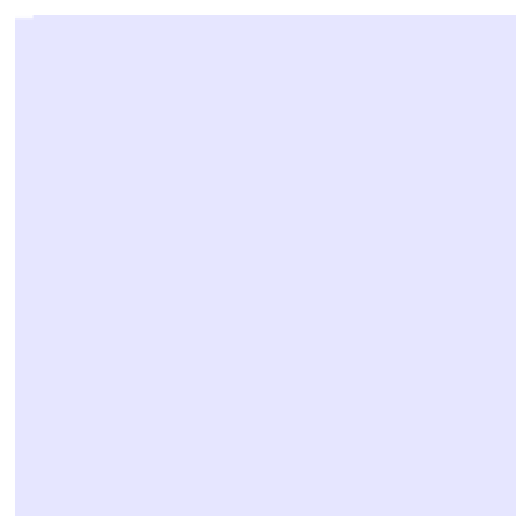}} & 
    \raisebox{-0.5\height}{\includegraphics[width=\b\textwidth]{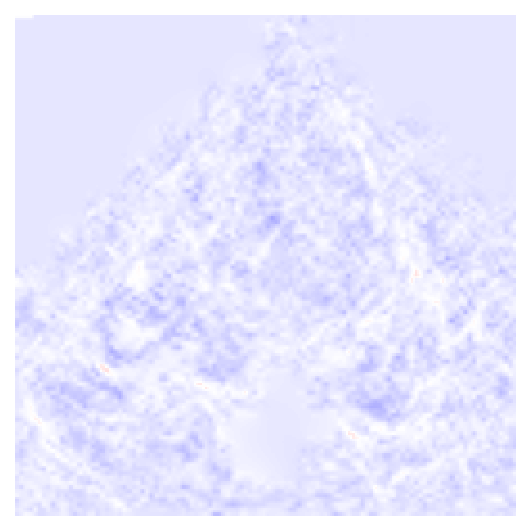}} & 
    \raisebox{-0.5\height}{\includegraphics[width=\b\textwidth]{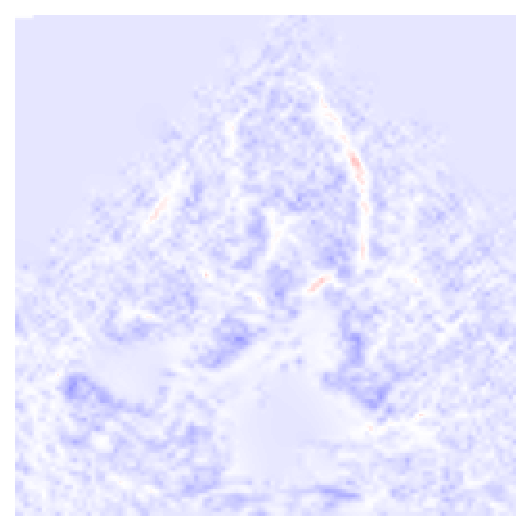}} &  
    \raisebox{-0.5\height}{\includegraphics[width=\b\textwidth]{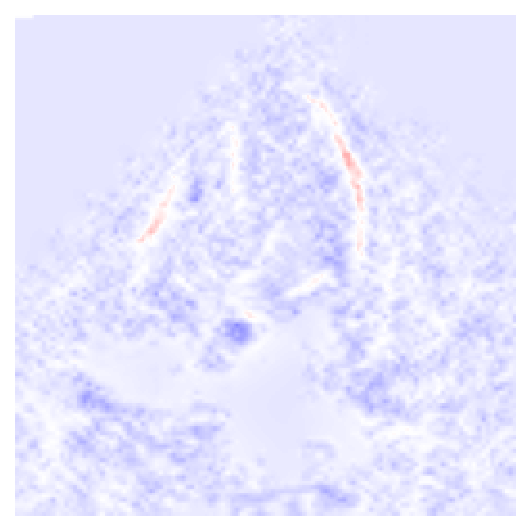}} &
    \raisebox{-0.5\height}{\includegraphics[width=\b\textwidth]{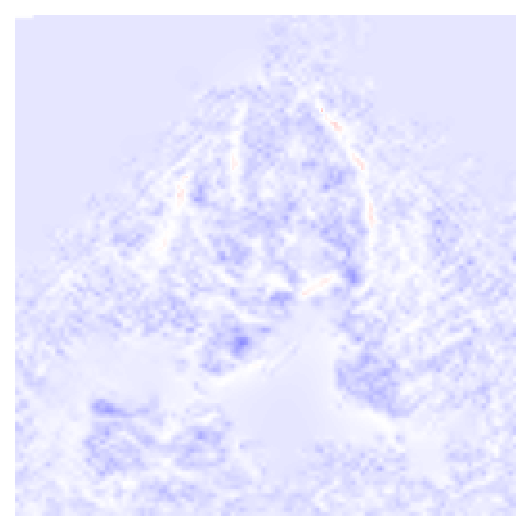}} & 
    \raisebox{-0.5\height}{\includegraphics[width=\b\textwidth]{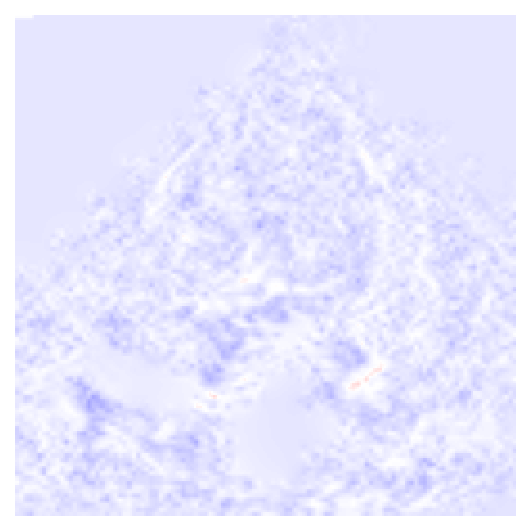}} & 
    \raisebox{-0.5\height}{\includegraphics[width=\b\textwidth]{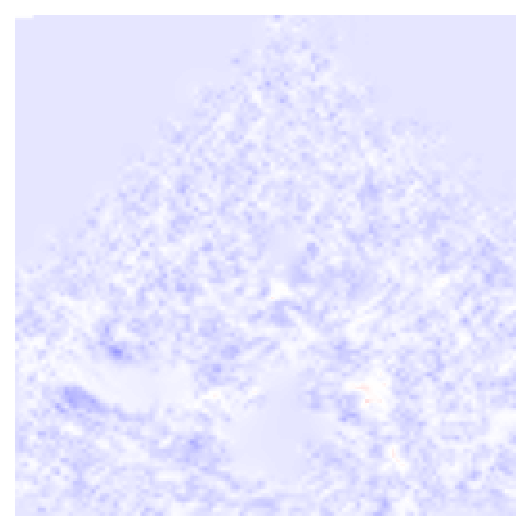}} & 
    \raisebox{-0.5\height}{\includegraphics[width=\b\textwidth]{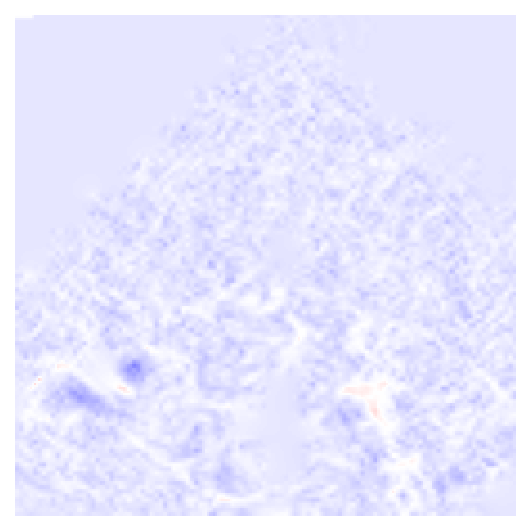}} 
    \\
    & & \multicolumn{8}{c}{\raisebox{-0.5\height}{\includegraphics[width=0.7\textwidth]{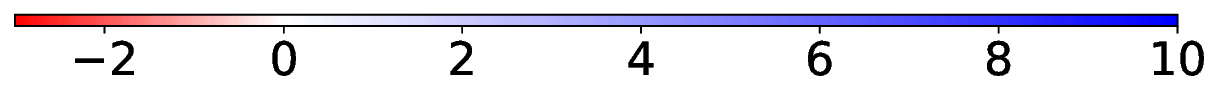}}} \\
    \bottomrule
  \end{tabular*}
  }
}
    \caption{Reconstructions and Jacobians for a single time sequence using our model and the Demons algorithm, We transfer a segmented region from time point, $t=1$ to the other in the sequence. In this example, the Dice coefficient between the human expert segmentation at $t=16$ and our estimate is $0.732$. The Dice coefficient using Demons algorithm is $0.715$.}
    \label{fig:dice_result}
\end{figure*}

In Figure~\ref{fig:dice_result}, we show the result of one sequence where we transform both the image and the human expert segmentation from one time step to the other. In this example the Dice coefficient between the other human expert segmentation at time $16$: our estimation is $0.732$, the Demons algorithm is $0.715$, and no registration is $0.583$.

\subsubsection{Latent analysis}
\begin{figure}[ht!]
    \centering
    \resizebox{1.0\linewidth}{!}{
\def\b{0.3}
\resizebox{\textwidth}{!}{
\begin{tabular*}{\textwidth}{m{1em}ccc}
    \toprule
     & $t=[1,\dots, 50]$ & $t=[1, 4, 7, \dots, 49]$ & $t=[1, \dots, 45]$ \\
    \toprule
    \rotatebox{90}{Smooth} & 
    \makecell{\includegraphics[width=\b\textwidth]{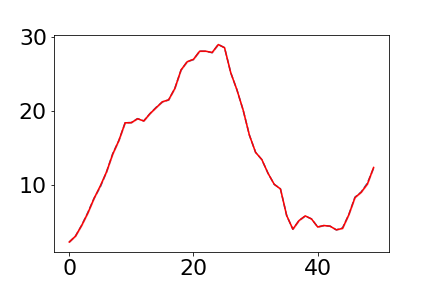}} & \makecell{\includegraphics[width=\b\textwidth]{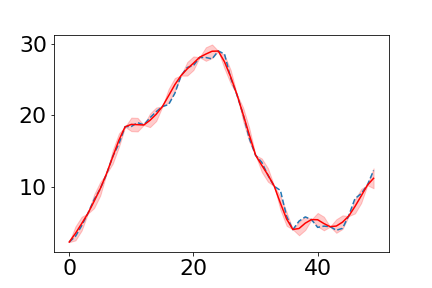}} & 
    \makecell{\includegraphics[width=\b\textwidth]{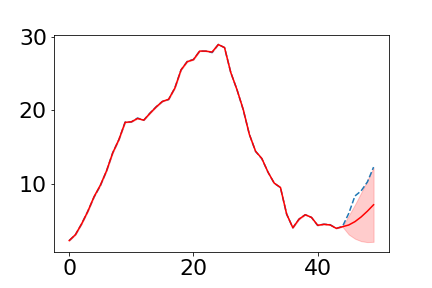}}  \\
    \rotatebox{90}{Filter} & 
    \makecell{\includegraphics[width=\b\textwidth]{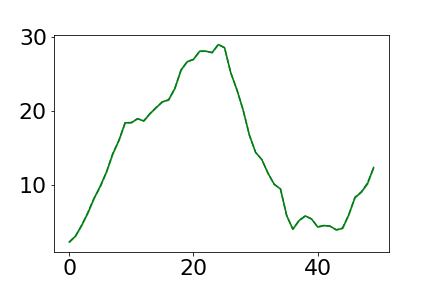}} & 
    \makecell{\includegraphics[width=\b\textwidth]{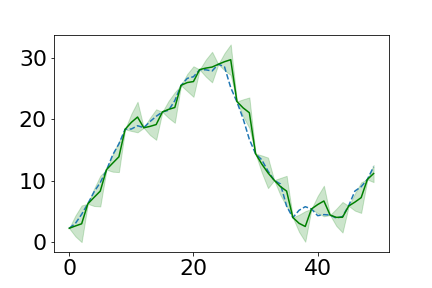}} & 
    \makecell{\includegraphics[width=\b\textwidth]{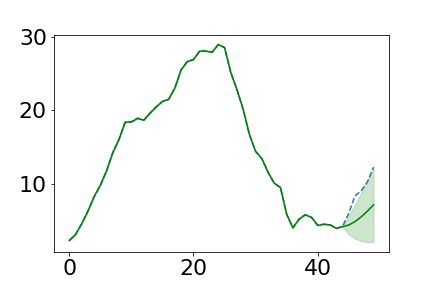}} \\
    \rotatebox{90}{Pred} & 
    \makecell{\includegraphics[width=\b\textwidth]{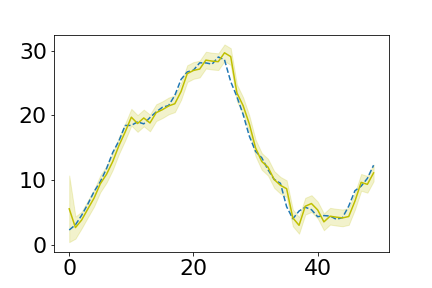}} & 
    \makecell{\includegraphics[width=\b\textwidth]{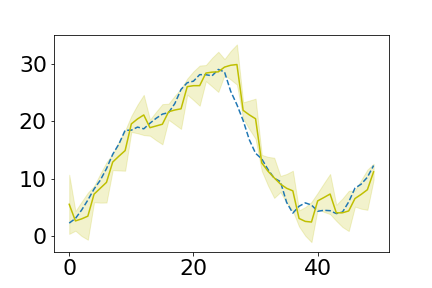}} & 
    \makecell{\includegraphics[width=\b\textwidth]{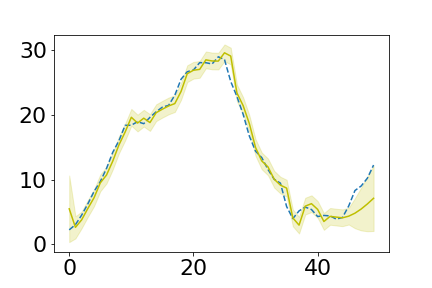}} \\
    \bottomrule
  \end{tabular*}
  }
}
    \caption{Distributions of the smooth, filtered, and one-step prediction distribution given observed data for one of the 16 dimensions in the latent observation space. The blue dashed lines are the observed latent variable for the entire sequence, the distributions are represented by their average (solid line), and standard deviation (shaded region).}
    \label{fig:latent}
\end{figure}

From the dynamics in the latent space, we can estimate the filtered, smooth, and predictive distributions. In Figure~\ref{fig:latent} we show those estimates for one of the latent space dimensions given three examples of input data: all observed values are known, imputation with every $3$th sample observed, and extrapolation for the last $5$. We here observed a continuous curve from where it is possible to impute and extrapolate for missing values.

\subsubsection{Extrapolating image sequence}
\begin{figure}[ht!]
    \centering
    \resizebox{1.0\linewidth}{!}{
\def\b{0.15}
\resizebox{\linewidth}{!}{
\begin{tabular*}{\textwidth}{m{1em}cccccc}
    \toprule
     & $t=1$ & $t=10$ & $t=20$ & $t=30$ & $t=40$ & $t=50$\\
    \toprule
    \rotatebox{90}{Images} & 
    \makecell{\includegraphics[width=\b\textwidth]{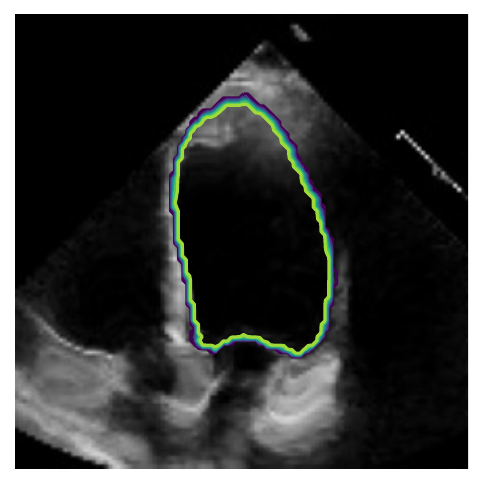}} & 
    \makecell{\includegraphics[width=\b\textwidth]{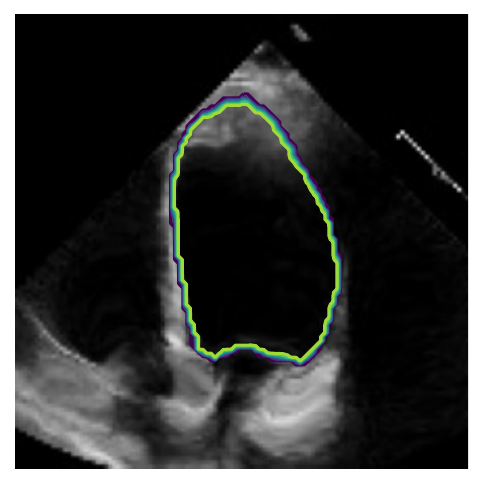}} &
    \makecell{\includegraphics[width=\b\textwidth]{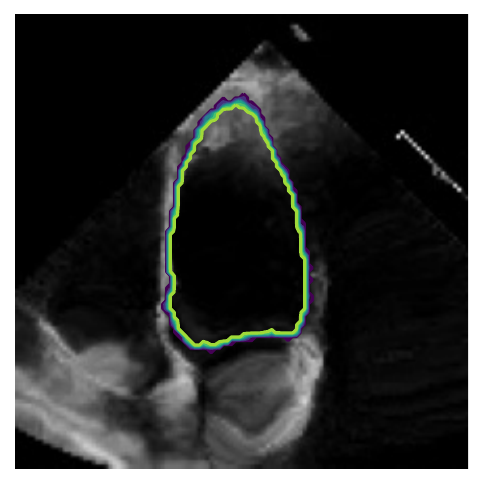}} &
    \makecell{\includegraphics[width=\b\textwidth]{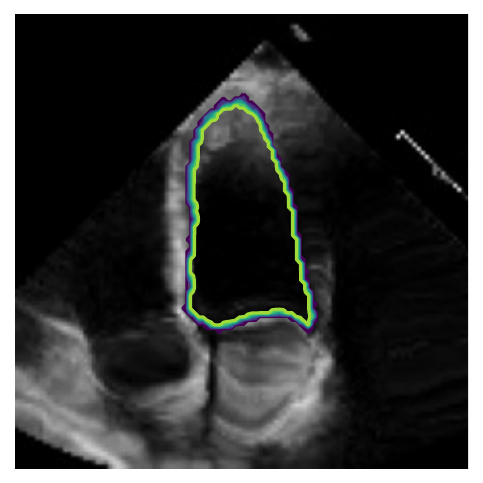}} &
    \makecell{\includegraphics[width=\b\textwidth]{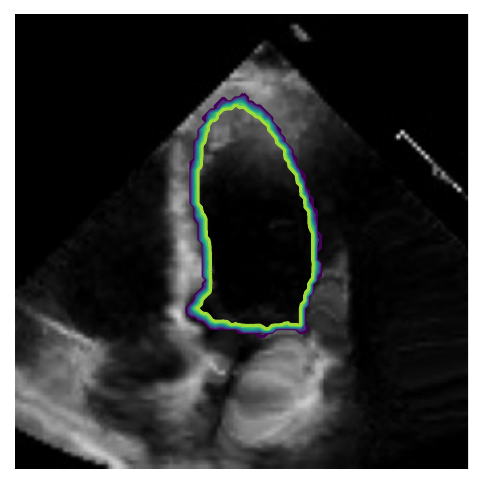}} &
    \makecell{\includegraphics[width=\b\textwidth]{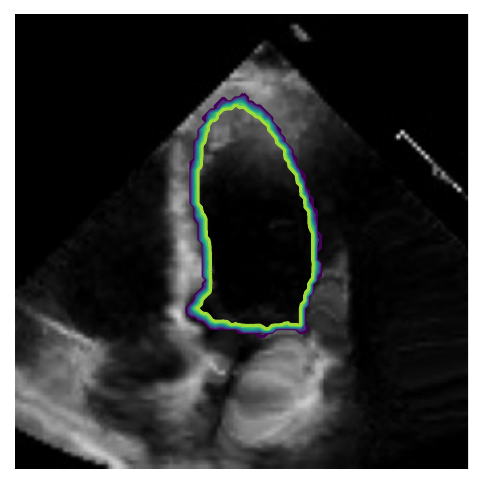}} \\
    \rotatebox{90}{$\varphi_t$} & 
    \makecell{\includegraphics[width=\b\textwidth]{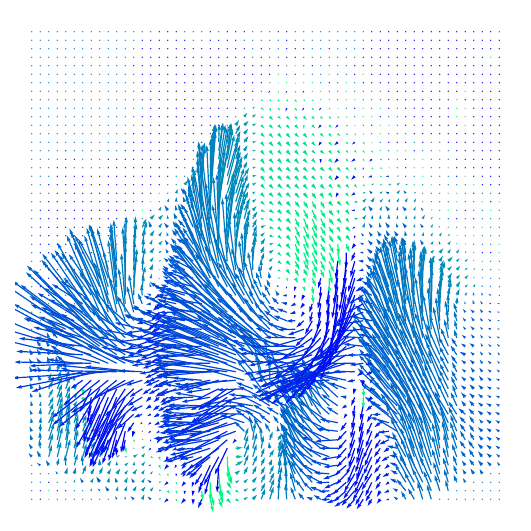}} & 
    \makecell{\includegraphics[width=\b\textwidth]{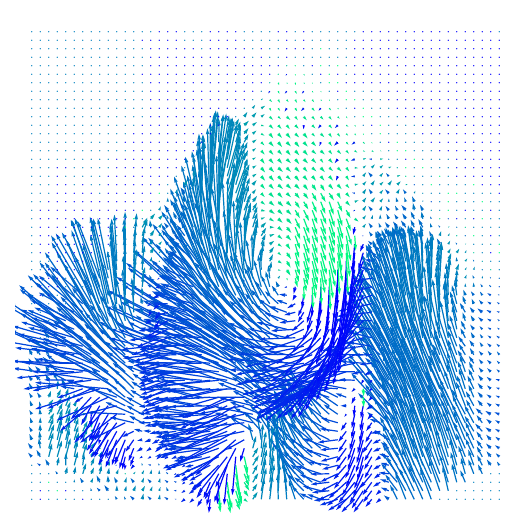}} &
    \makecell{\includegraphics[width=\b\textwidth]{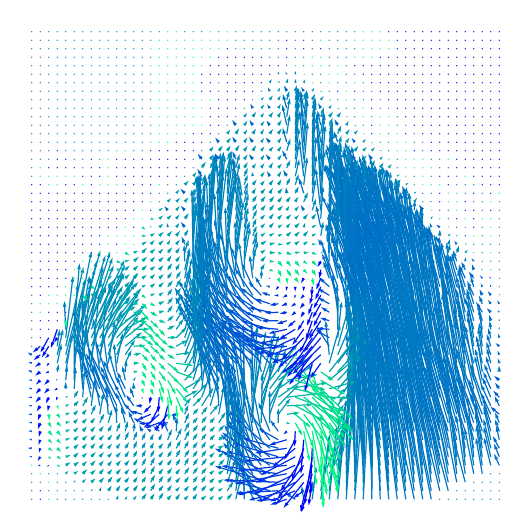}} &
    \makecell{\includegraphics[width=\b\textwidth]{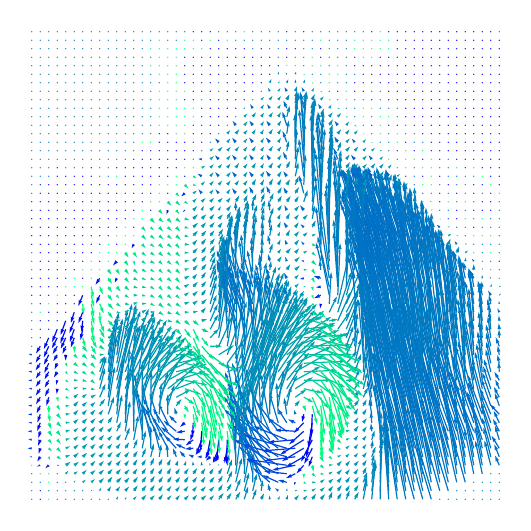}} &
    \makecell{\includegraphics[width=\b\textwidth]{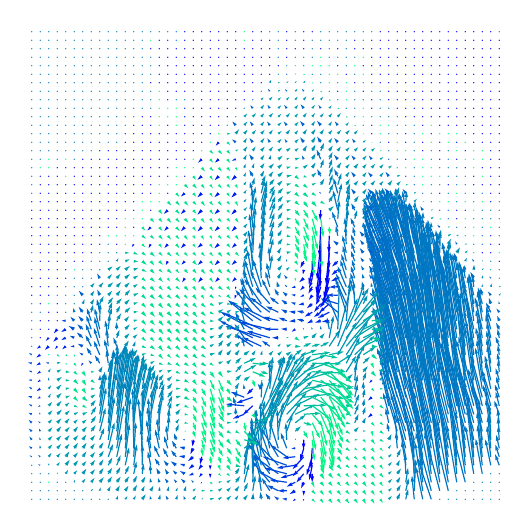}} &
    \makecell{\includegraphics[width=\b\textwidth]{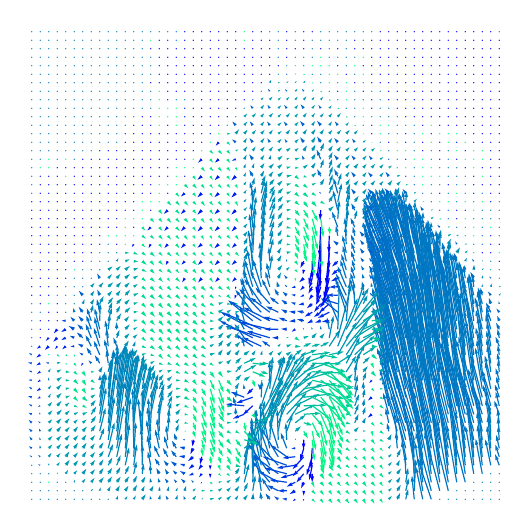}} \\
    \rotatebox{90}{$|J_\varphi|$} & 
    \makecell{\includegraphics[width=\b\textwidth]{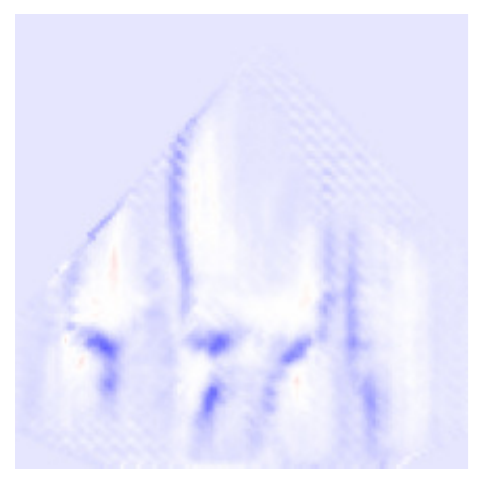}} & 
    \makecell{\includegraphics[width=\b\textwidth]{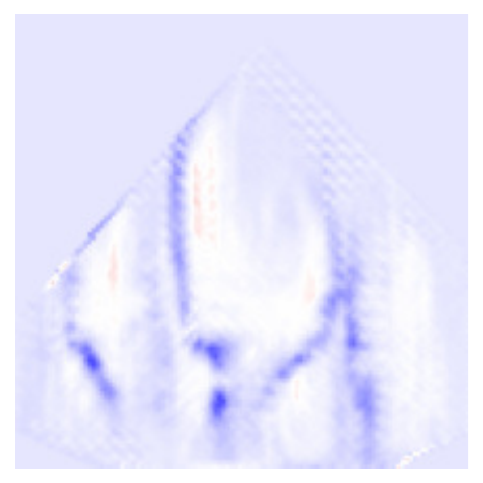}} &
    \makecell{\includegraphics[width=\b\textwidth]{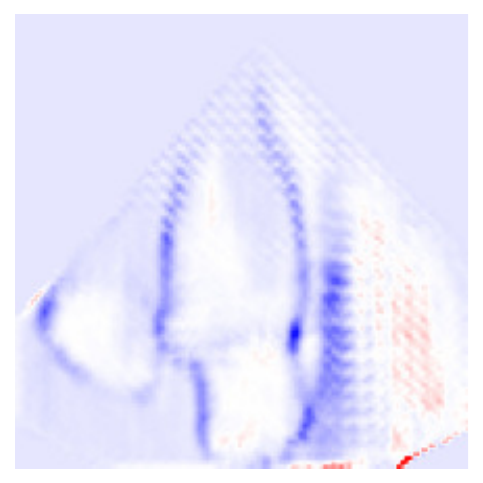}} &
    \makecell{\includegraphics[width=\b\textwidth]{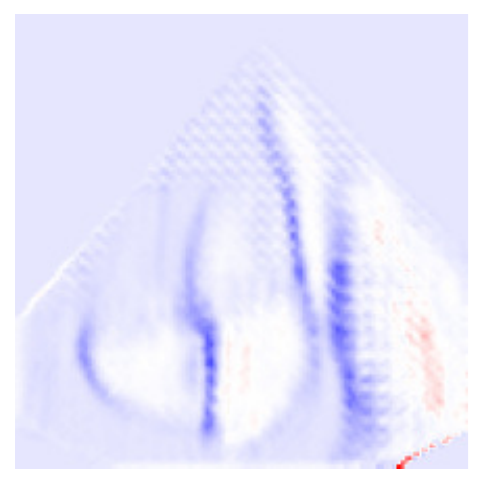}} &
    \makecell{\includegraphics[width=\b\textwidth]{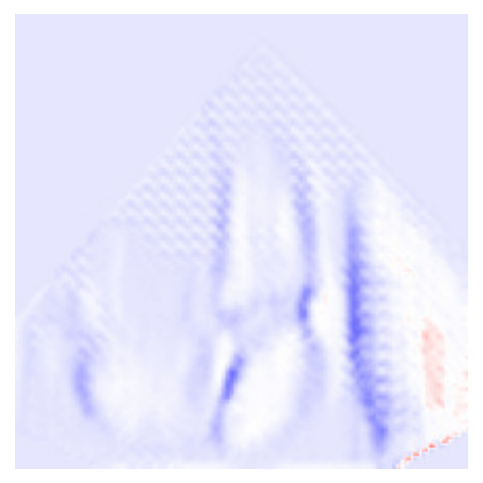}} &
    \makecell{\includegraphics[width=\b\textwidth]{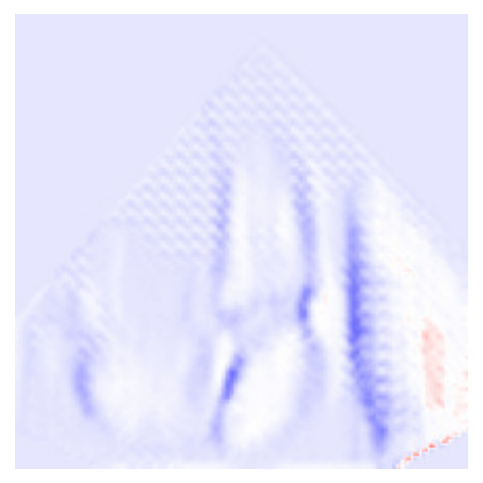}} \\
    & \multicolumn{6}{c}{\makecell{\includegraphics[width=0.7\textwidth]{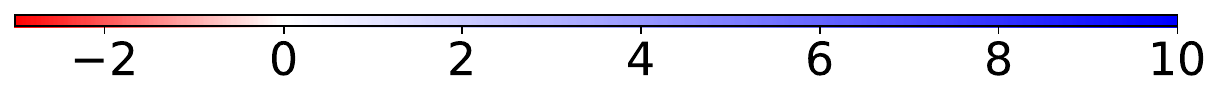}}} \\
    \bottomrule
  \end{tabular*}
  }
}
    \caption{Generated transformations by sampling an initial position in the latent space. Here we show the transformed moving image (top), the displacement field (middle) and the Jacobian of the generated data (bottom).}
    \label{fig:gen}
\end{figure}

The model is generative --- given a moving image, $y_M$ we can generate a displacement field by sampling from the initial prior in the latent space and propagating this sample forward in time. Figure~\ref{fig:gen} illustrates the generated result for one sample.

\section{Discussion}
We have described an unsupervised method for extracting latent linear dynamics directly from a medical image time series and show how we can reconstruct the displacement field from the lower-dimensional latent linear dynamic system. With a similar performance in Dice score, our model outperformed the conventional method in speed. 

We also illustrate how we can impute missing samples and extrapolate forward in time. Extrapolation support is beneficial for real-time systems when action must be taken given observed data. Good estimates of future states may be necessary to compensate for the system latency. This has e.g. been investigated for the case of real-time adaptive radiotherapy where latency in the motion feedback loop has been seen to cause a decrease in the quality of the treatment~\cite{bedford2015effect}.

However, we also verified implausible transformation vectors, i.e., Jacobian determinants with non-positive values. Those values mostly appear in regions of the images where the intensity is zero ($67 \%$). This is not surprising since we do not handle the regularization of the displacement field directly in the model. We showed that a low-pass Gaussian filter reduces the number of non-negative Jacobian determinants without significantly affecting performance. We estimate the transformation likelihood as an independent Gaussian with fixed variance which is probably a too strong assumption. More likely, the intensities of nearby pixels are correlated, which could be captured for instance by having a Laplacian structured precision matrix. We believe a richer explanation of the transformation likelihood could address this better, and also provide a better uncertainty estimation of the image transformation. Another approach to regularize the transformation could be to estimate a sparse representation of the displacement field and use interpolation techniques~\cite{gunnarsson2020registration}, like B-splines~\cite{rueckert1999nonrigid} or thin plate splines~\cite{bookstein1991thin} to estimate the dense representation. Furthermore, diffeomorphic deformations can enforced by applying exponential layers~\cite{krebs2019learning} to the network.

In our experiment, we estimated the 2D dynamics based on 2D echocardiogram time series. As further work, we want to extend this to the complete 3D motion.

\section*{Acknowledgement}
This research was supported by \emph{Wallenberg AI, Autonomous Systems and Software Program (WASP)} funded by Knut and Alice Wallenberg Foundation, and the Swedish Foundation for Strategic Research grant SM19-0029.
\printbibliography

\end{document}